\documentclass[preprint,showpacs,prb,superscriptaddress]{revtex4-1}

\usepackage{amsmath}
\usepackage{graphicx}
\usepackage{subfigure}

\newcommand{\beq}{\begin{equation}}
\newcommand{\eeq}{\end{equation}}
\newcommand{\inlinegraph}[1]{\vcenter{\hbox{\includegraphics{#1}}}}
\DeclareMathOperator{\sgn}{sgn}

\begin{document}

\title{Loops, sign structures and emergent Fermi statistics in three-dimensional quantum dimer models}

\author{Vsevolod Ivanov}\thanks{Present Address: Department of Physics, California Institute of Technology, Pasadena, CA 91125, USA}
\affiliation{Department of Physics, Massachusetts Institute of Technology, Cambridge, MA 02139}
\author{Yang Qi}
\affiliation{Institute of Advanced Study, Tsinghua University, Beijing 100084, China}
\author{Liang Fu}
\affiliation{Department of Physics, Massachusetts Institute of Technology, Cambridge, MA 02139}

\begin{abstract}
  We introduce and study three-dimensional quantum dimer models with
  positive resonance terms.  We demonstrate that their ground state
  wave functions exhibit a nonlocal sign structure that can be exactly
  formulated in terms of loops, and as a direct consequence, monomer
  excitations obey Fermi statistics.  The sign structure and Fermi
  statistics in these ``signful'' quantum dimer models can be
  naturally described by a parton construction, which becomes exact at
  the solvable point.
\end{abstract}

\pacs{75.10.Kt, 05.30.-d, 71.10.Pm}

\maketitle

The Rokhsar-Kivelson (RK) quantum dimer model was originally
introduced\cite{rk} to describe the short-range
resonating-valence-bond (RVB) state in two dimensions\cite{anderson}.
Later it was discovered that quantum dimer models provide particularly
simple and elegant realizations of topological phases of matter,
including a two-dimensional (2D) gapped phase with $Z_2$ topological
order\cite{sondhi}, and a three-dimensional (3D) Coulomb phase
described by an emergent Maxwell electrodynamics\cite{hermele, huse}.
Both phases possess a fractional quasiparticle called a monomer, a
deconfined pointlike excitation that carries half of the $U(1)$ charge
of a dimer.

The statistics of monomers was a subject of considerable attention and
debate in early studies\cite{krs, haldane} and of continuing interest
recently\cite{poilblanc}. It was eventually settled\cite{kivelson,
  read} that the statistics of monomers in two-dimensional quantum
dimer models cannot be assigned in a universal way, because statistics
can be altered by attaching a $\pi$ flux (vison\cite{senthil}) to a
monomer. On the other hand, a boson cannot be changed into a fermion
by $\pi$-flux binding in three dimensions, because particles and flux
lines are objects of different dimensions. This leaves open the
possibility of monomers with Fermi statistics in 3D quantum dimer
models, which is the subject of this work.  We note that Fermi
statistics also arises in other 3D boson models with an emergent $Z_2$
gauge field\cite{lw, vishwanath}.

We introduce and study quantum dimer models in 3D non-bipartite
lattices with {\it positive} resonance terms, in contrast to terms
with negative coefficients in the original RK model\cite{rk}.  We find
a ``twisted'' $Z_2$ topological phase in which monomers are deconfined
and obey Fermi statistics. The Fermi statistics arises from a nonlocal
sign structure of the ground state wave function, which is specified
through loops in the transition graph by an exact sign rule.  We
provide a parton construction for these ``signful'' quantum dimer
models, which yields the exact ground state dimer wave function and
naturally explains the emergent Fermi statistics of monomers.
Finally, we discuss the implications of our study for doped quantum
dimer models and resonating-valence-bond (RVB) states in three
dimensions.

\begin{figure}[ht]
\centering
\includegraphics[width=.45\textwidth]{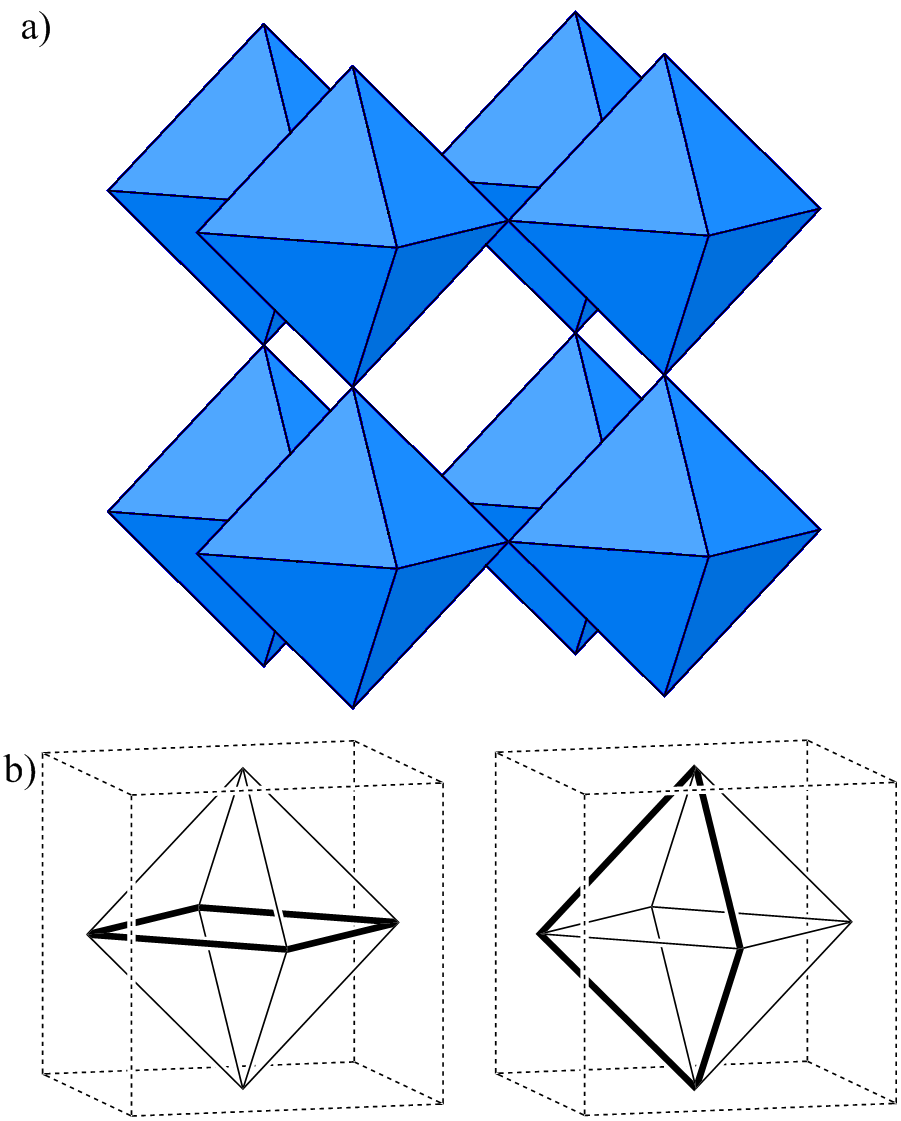}
\caption{(a) Eight cubic unit cells of the CSO lattice. (b) Two types
  of length-4 loops on the CSO lattice. Left: square loop, right:
  bent-square loop.}
\label{CSOloops}
\end{figure}

While our main results generically apply to 3D quantum dimer models on
non-bipartite lattices, we illustrate the essential physics using a 3D
lattice made of corner-sharing octahedra (CSO), shown in
Fig.~\ref{CSOloops}(a). Each site is touched by exactly one dimer
occupying one of the eight nearest-neighbor bonds. The quantum dimer
model Hamiltonian $H$ consists of resonance moves that locally flip a
dimer configuration, and potential terms that give an energy cost for
every flippable configuration.

For simplicity we only consider shortest resonance loops of length
four.  There exist two types of length-four resonance loops, having
the shape of a square and a bent square respectively [see
Fig.~\ref{CSOloops}(b)].  $H$ is thus given by
\begin{equation}
  \label{H}
  \begin{split}
  H = & J_1 \left(\left|\inlinegraph{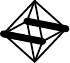}\right\rangle
   \left\langle\inlinegraph{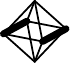}\right|+
   \left|\inlinegraph{jk2}\right\rangle
   \left\langle\inlinegraph{jk1}\right|\right)\\
 & + K_1 \left(\left|\inlinegraph{jk1}\right\rangle
   \left\langle\inlinegraph{jk1}\right|+
   \left|\inlinegraph{jk2}\right\rangle
   \left\langle\inlinegraph{jk2}\right|\right)\\
 &+ J_2 \left(\left|\inlinegraph{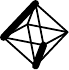}\right\rangle
   \left\langle\inlinegraph{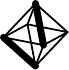}\right|+
   \left|\inlinegraph{jk4}\right\rangle
   \left\langle\inlinegraph{jk3}\right|\right)\\
 & + K_2 \left(\left|\inlinegraph{jk3}\right\rangle
   \left\langle\inlinegraph{jk3}\right|+
   \left|\inlinegraph{jk4}\right\rangle
   \left\langle\inlinegraph{jk4}\right|\right)\\
  &+ \cdots    
  \end{split}
\end{equation}
Here $\cdots$ denotes similar terms for resonance loops of longer
length, whose form will be discussed later.

When $ |J_i| = K_i>0$, $H$ can be written as a sum of positive
semidefinite projection operators:
\begin{equation}
  \label{eq:Hproj}
  \begin{split}
  H=&K_1\left(\left|\inlinegraph{jk1}\right>+
    \eta_1\left|\inlinegraph{jk2}\right>\right)
  \left(\left<\inlinegraph{jk1}\right|+
    \eta_1\left<\inlinegraph{jk2}\right|\right)\\
  &+K_2\left(\left|\inlinegraph{jk3}\right>+
    \eta_2\left|\inlinegraph{jk4}\right>\right)
  \left(\left<\inlinegraph{jk3}\right|+
    \eta_2\left<\inlinegraph{jk4}\right|\right)
  \end{split}
\end{equation}
where $\eta_i = J_i/ K_i =\pm 1$.  $\eta_1=\eta_2=-1$ corresponds to
the original RK solvable point with negative resonance terms. In this
case, the ground state is an equal {\it amplitude} superposition of
all dimer coverings. Such a dimer-liquid state in a 3D CSO lattice is
expected to represent a gapped $Z_2$ topological phase similar to the
one in a face-centered-cubic lattice\cite{huse}.  Since this RK wave
function is everywhere positive, monomer excitations above the ground
state are necessarily bosons.

Here we study quantum dimer models with {\it positive} resonance
terms: $J_1, J_2 >0$.  In this case, ground states are ``signful'',
taking both positive and negative values for different dimer
coverings.  Such a nontrivial sign structure is a prerequisite to the
emergence of fermionic monomer excitations.  Following the spirit of
the RK approach, we first construct a signful dimer wave function and
then demonstrate that this wave function is the zero-energy ground
state of $H$ at a generalized RK point $J_i = K_i >0$.

\begin{figure}[ht]
\centering
\includegraphics[width=.3\textwidth]{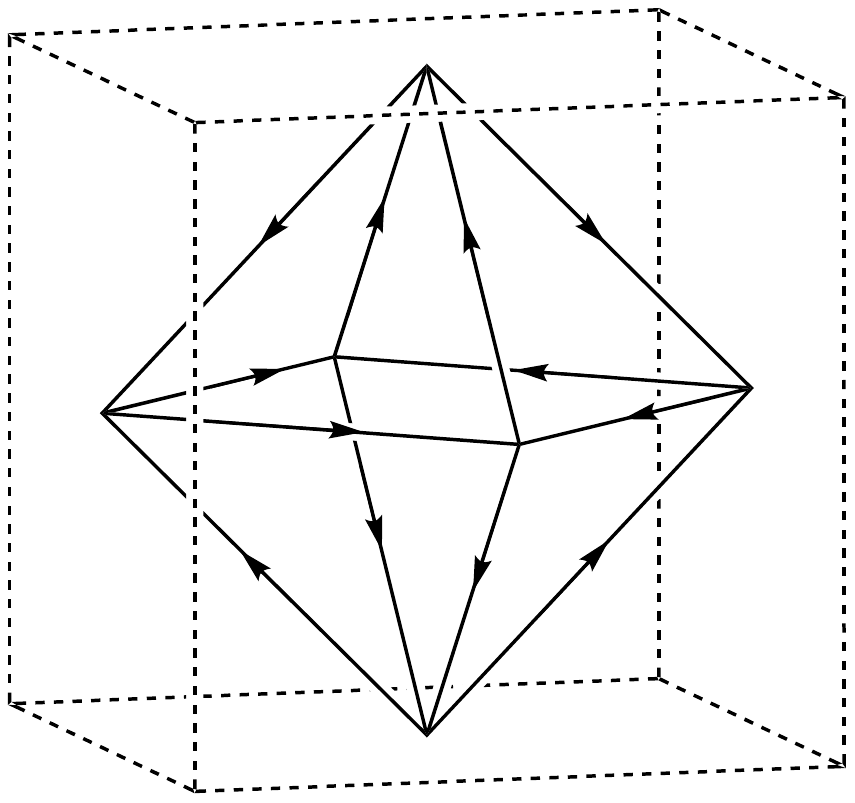}\\
\caption{Arrow pattern for CSO lattice.}
\label{CSOarrows}
\end{figure}

\section{Loops and sign structures.}
\label{sec:loops-sign-struct}

The dimer wave function we constructed is an equal {\it weight}
superposition of all dimer coverings $\{ D_a\}$ with $a=1,\ldots,N_D$
with a {\it sign} structure: \beq \Psi(D_a)= \frac{1}{\sqrt{N_D}}
s(D_a). \label{psi} \eeq The key ingredient for the emergent Fermi
statistics is the form of $s(D_a)$, which we now define. First, we
create a transition graph $G_{1a}$ by superimposing $D_a$ and an
arbitrary reference dimer covering $D_1$.  The transition graph
consists of non-intersecting even-length loops. Since dimer coverings
and transition loops are in one-to-one correspondence, we define
$s(D_a) \equiv s(G_{1a})$ as a function of transition loops.

The sign of transition loops is specified by endowing every nearest-neighbor bond with an arrow. The arrow pattern we choose for the CSO lattice is shown in Fig.~\ref{CSOarrows}. Now we define $s(G_{ab})$ in terms of loops and arrows:
\beq
s (G_{ab}) = (-1)^{N(G_{ab}) + W(G_{ab})}, \label{s}
\eeq    
where  $N(G_{ab})$ is the total number of loops in $G_{ab}$, and $W(G_{ab})$ is the total number of ``wrong-way'' arrows when all loops are 
traversed unidirectionally. 

The sign structure (\ref{s}) we introduced for the quantum dimer wave
function (\ref{psi}) is a central result of this work.  It has several
important properties. First, because all loops in the transition graph
have an even length, $(-1)^{W(G_{ab})}$ is independent of the
direction of traverse, as it should be for loops in three dimensions.
Second, $s(G_{ab})$ is multiplicative under the composition of
transition loops. For three different dimer coverings $D_a$, $D_b$ and
$D_c$,
\begin{equation}
  \label{eq:smul}
  s(G_{ac}) = s(G_{ab}) s(G_{bc}).   
\end{equation}
This is obvious when $G_{ab}$ and $G_{bc}$ do not overlap, and is also
true when they do. For example, when $G_{ab}$ and $G_{bc}$ overlap on
one bond, two loops combine into one in $G_{ac}$, which reduces both
the total number of loops and the number of wrong-way arrows by
one. This result is proved for general cases in
Appendix~\ref{sec:addit-exampl-signs} and an alternative proof based
on parton construction is given in
Sec.~\ref{sec:parton-construct}. Equation~(\ref{eq:smul}) guarantees
that the dimer wave function defined using the sign structure
$s(G_{ab})$ does not depend on the choice of the reference dimer
covering, up to an overall factor of $-1$.  Last but not the least,
the arrow pattern in Fig.~\ref{CSOarrows} guarantees that $\Psi(D_a)$
is invariant under all symmetry transformations of the CSO lattice,
despite that the arrow pattern itself is not.

To justify the last point, we consider the symmetry transformation of the arrow pattern shown in Fig.~\ref{CSOarrows}. The symmetry of the CSO lattice can be represented by the $O_h$ point group, along with translation symmetry. We now check how the arrows transform under the symmetry operations of $O_h$, with 
$+1$ signifying that all arrows are unchanged, and $-1$ signifying that all arrows reverse:
\begin{center}
	\begin{tabular}{ l | c c c c c c c c c c}
			$O_h$ & $E$ & $8C_3$ & $6C_2$ & $6C_4$ & $3C^2_4$ & $i$ & $6S_4$ & $8S_6$ & $3\sigma_h$ & $6\sigma_d$\\ \hline
			$A_{2g}$ & $+1$ & $+1$ & $-1$ & $-1$ & $+1$ & $+1$ & $-1$ & $+1$ & $+1$ & $-1$\\ 
	\end{tabular}
\end{center}
This arrow pattern behaves as the representation $A_{2g}$ of $O_h$
point group. For a given transition graph, the sign of a particular
loop given by Eq.~(\ref{s}) does not change under the transformation
of reversing all arrows, as each loop contains even number of
dimers. Therefore the wave function defined in Eq.~(\ref{psi}) is
invariant under symmetry.

According to the arrow pattern, the signs $s(G_{ab})$ of the two
shortest-length transition loops, square and bent square, are $-1$. It
then follows that the dimer wave function $|\Psi\rangle$ is a
zero-energy ground state of the quantum dimer model Hamiltonian
(\ref{H}) at the generalized RK point $J_i=K_i$. This completes our
construction of solvable 3D quantum dimer models with a nontrivial
sign structure.

It is worth pointing out that, as is common with quantum dimer models
at RK point, the Hamiltonian (\ref{H}) has other zero-energy grounds
states in addition to $\Psi(D_a)$, such as ``dead'' dimer coverings
that are completely non-flippable. Also, similar to quantum dimer
models on other 3D lattices\cite{Sikora2011}, we found that with only
resonance terms on length-four loops the Hamiltonian does not connect
all coverings in a given topological sector. By including additional
resonance terms involving longer loops and choosing their signs
according to the arrow pattern, one can increase the connectivity of
dimer coverings, although the issue of ergodicity in 3D quantum dimer
models is beyond the scope of this paper. On a positive side, we
expect the dimer liquid state (\ref{psi}) to be stabilized in an
extended regime $|J_i|>K_i$, as found for other non-bipartite
lattices\cite{sondhi, huse}. This issue can be studied numerically
using Monte Carlo methods\cite{Sikora2011, qmcnote}.

  
\section{Statistics of monomers.}
\label{sec:stat-mono}

We now demonstrate that the nontrivial sign structure of the ground
state (\ref{s}) directly gives rise to Fermi statistics of the
monomer, a pointlike excitation associated with a site that is not
touched by any dimer. Since a missing dimer can break up into two
monomers, each monomer is a charge-$1/2$ fractional quasiparticle,
which has a finite excitation energy associated with breaking the rule
of one dimer per site.

Following the approach of Ref.~\onlinecite{krs}, we determine the
statistics by adiabatically transporting two monomers along a path
which exchanges their positions, and examining the resulting change in
the phase of the many-body wave function. To implement the exchange,
we introduce dimer move terms $H_T$ into the quantum dimer model,
which enables a monomer to hop by exchanging with a nearby
dimer. Without losing generality we consider the following form of
$H_T$,
\begin{equation}
  \label{eq:HT}
  H_T=\lambda\sum\left(\left|\inlinegraph{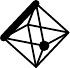}\right>
  \left<\inlinegraph{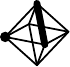}\right|+\text{h. c.}\right),
\end{equation}
where the sum runs over all symmetry-related moves that transport a
monomer to the nearest-neighbor site (see Fig.~3). 
Here we consider only nearest-neighbor hoppings, but a general proof
of Fermi statistics for general hopping terms is given in
Sec.~\ref{sec:parton-construct}.

We assume that $\lambda$ is small and moves the monomer
adiabatically. In the limit of $\lambda\rightarrow0$, the eigenstates
of the system are the dimer wave functions with monomers at fixed
locations. Particularly the dimer wave function with a monomer at site
$i$, denoted as $|\Psi_i\rangle$, is a superposition of dimer
coverings that has no dimer connecting to site $i$, with the sign rule
in Eq.~(\ref{s})\cite{monomernote}. Under the adiabatic assumption,
each time $H_T$ is applied to a monomer state to move a monomer from
$i$ to $j$, it takes dimer coverings in $|\Psi_i\rangle$ and converts
them to dimer coverings in $|\Psi_j\rangle$, and then the state of
converted dimer coverings relaxes into the eigenstate $|\Psi_j\rangle$
before the monomer is moved again. Therefore the relative sign of the
wave functions before and after the hopping is determined by projecting
$H_T$ onto $|\Psi_i\rangle$ and $|\Psi_j\rangle$ as the following,
\begin{equation}
  \label{eq:tij}
  t_{ij} = \langle\Psi_j|H_T|\Psi_i\rangle.
\end{equation}

\begin{figure}[htbp]
  \centering
  \includegraphics{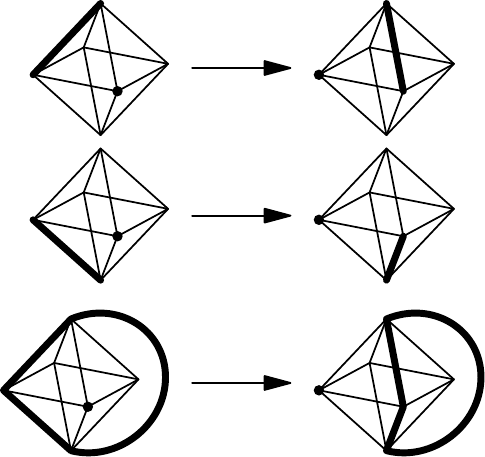}
  \caption{Two different ways to move a monomer. The first and second
    rows show two different ways to move a monomer, and the last row
    shows the transition graph obtained by compositing the two dimer
    coverings in the same column. The curvy segment shows the part of
    the transition loop through the rest of the lattice, which is
    identical before and after the monomer move.}
  \label{fig:exchange}
\end{figure}

There are different terms in $H_T$ that contribute to $t_{ij}$, which
correspond to different ways to exchange a monomer and a dimer to move
the former from $i$ to $j$, as illustrated in
Fig.~\ref{fig:exchange}. Here we show that these different terms
contribute with the same sign to $t_{ij}$ in Eq.~\eqref{eq:tij}. In
Fig.~\ref{fig:exchange} it is illustrated that there are two ways to
move a monomer from $i$ to $j$ by exchanging it with two different
dimers, as shown on the first and second row. The relative phase
between the two different dimer coverings in the initial and final
states is determined by the transition loop on the bottom. The two
transition loops in the initial and final states differ by moving two
dimers from one side of a bent-square loop to the other
side. Therefore the two transition loops have the same sign according
to Eq.~(\ref{s}), as the arrow pattern shown in Fig.~\ref{CSOarrows}
has an even number of wrong-way arrows in any four-bond loop. This
shows that different terms in Eq.~\eqref{eq:tij} contribute to
$t_{ij}$ with the same sign. Consequently the sign of $t_{ij}$ can be
determined from any two dimer coverings in $|\Psi_i\rangle$ and
$|\Psi_j\rangle$ that can be connected by $H_T$. Generally, this
property holds for any arrow pattern with an even number of wrong-way
arrows in four-bond loops, and a proof will be given in
Sec.~\ref{sec:parton-construct}.

\begin{figure}[ht]
\centering
\includegraphics[width=.45\textwidth]{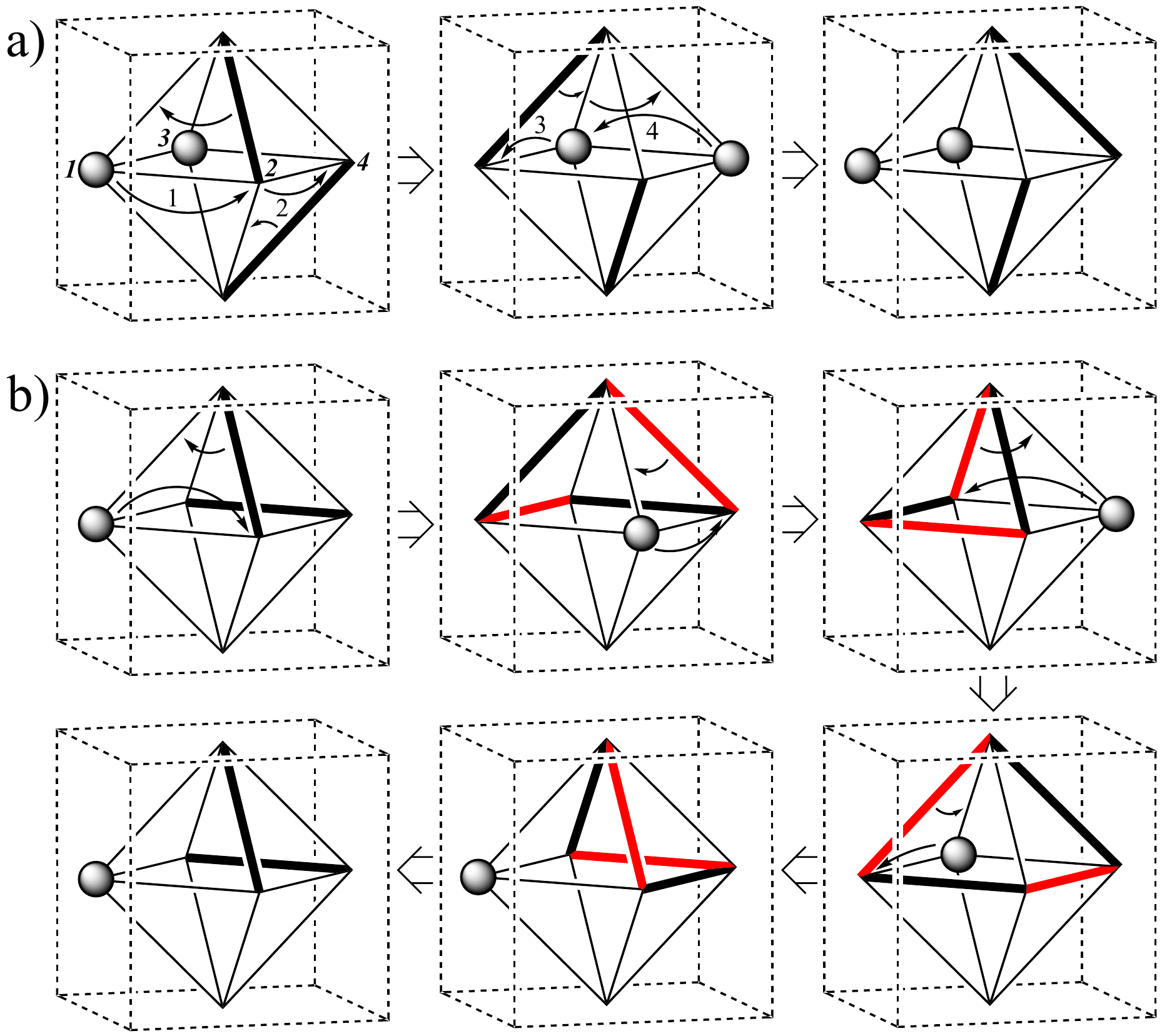}\\
\caption{(a) The two holes are exchanged using nearest-neighbor hops
  defined by numbered arrows. Possible hole positions are marked in
  italics. Dimers are flipped accordingly. (b) We move a single hole
  along the same path to find the flux, changing our reference tiling
  using dimer flips along bent-square loops when necessary. When such
  a dimer flip is performed, the resulting new dimer tiling is marked
  in red. In total, 4 bent square loops are used, giving a +1 sign. }
\label{NNexchange}
\end{figure}

Since quasiparticle statistics is path independent, we consider a
simple process of exchanging two monomers at initial positions $1$ and
$3$ via four nearest-neighbor monomer hops in the following sequence:
$1 \rightarrow 2$, $2\rightarrow 4$, $3\rightarrow 1$ and
$4\rightarrow 3$ (see Fig.~\ref{NNexchange}).  After the exchange, the
many-body state of the system comes back to itself, up to a possible
phase change $\theta_{ex}$. Since the dimer wave functions are real,
$\theta_{ex}$ is either $0$ or $\pi$. $\theta_{ex}$ can be determined
from the product of transition amplitudes at every stage of the
exchange process\cite{lw}
\begin{equation} 
  \label{te} 
  t_{ex} = t_{12} t_{24} t_{31} t_{43}.
\end{equation}
The sign of $t_{ex}$ is unambiguous (independent of the gauge choice
of dimer wave functions), and determines $\theta_{ex}$: $\theta_{ex}=0$
for $t_{ex}>0$ and $\pi$ for $t_{ex}<0$.

To evaluate (\ref{te}), we use a natural sign convention for the
many-body states $|\Psi_{ij}\rangle$ where $i, j$ label the positions
of the two monomers, such that their amplitudes on the first four
reference dimer coverings are positive:
\[\Psi_{13}(D_a)=\Psi_{23}(D_b) =\Psi_{34}(D_c) =\Psi_{14}(D_d) >0,\]
where the sequence of reference dimer coverings $D_a\rightarrow D_b
\rightarrow D_c \rightarrow D_d \rightarrow D_e$ are the ones that
follow the sequence of dimer moves as illustrated in
Fig.~\ref{NNexchange}.  Under this sign convention, $t_{12}, t_{24}$,
$t_{31}$ have the same sign.  Furthermore, since $D_e$ and $D_a$ only
differ by a resonance loop on a bent square, the sign structure
(\ref{s}) dictates $\Psi_{13}(D_e) = - \Psi_{13}(D_a)$. Comparing with
the convention $\Psi_{13}(D_a)>0$, this implies that $t_{43}$ have an
opposite sign from the rest of the $t$'s, which makes $t_{ex}$
negative. Therefore we conclude that the phase change after the
monomer exchange is $\theta_{ex}=\pi$.
  
$\theta_{ex}$ is given by the sum of the statistical angle $\theta_s$,
which is $0$ for bosons and $\pi$ for fermions, {\it and} the background
flux $\phi$ passing through the exchange paths of two particles that
join into a loop $\Gamma$.  Therefore to determine the quasiparticle
statistics requires the knowledge of $\phi$ in addition to
$\theta_{ex}$.  For a real wave function, $\phi$ can only be $0$ or
$\pi$.  $\phi$ can be determined from the phase change after
transporting a {\it single} monomer around the loop $\Gamma$; here
$\Gamma$ is the boundary of a single square plaquette. Thus we now use
the same set of dimer move operators $H_T$ and carry out a procedure
similarly as before to compute $\phi$, with two important
differences. First, the order in which the hopping terms are applied
must change in order to form a continuous path, which is reflected by
the new ordering of transition amplitudes:
\begin{equation} 
  t_{ex} = t_{12} t_{24} t_{43} t_{31}.
\end{equation}
Second, each hole hop must be followed by a change in the reference
dimer tiling, in order to guarantee that there exists a dimer-hole
configuration that allows the hopping operation to be applied. The
sequence of reference dimer tilings is $D_a\rightarrow D_b\Rightarrow
D_{b'}\rightarrow D_c \Rightarrow D_{c'}\rightarrow D_d \Rightarrow
D_{d'}\rightarrow D_e\Rightarrow D_{e'}$ where $\rightarrow$
represents a hole hop, while $\Rightarrow$ represents a change of
reference. This sequence of operations is shown in
Fig.~\ref{NNexchange}(b), with each $D_{i'}$ after a change of reference
marked in red. We note that $D_a = D_{e'}$, and that the final change
of reference $D_e\Rightarrow D_{e'}$ is carried out for the sake of
convenience, since it allows us to refer to all transition loops as
changes of reference. In this case, each change of reference
corresponds to a bent-square loop, and since we perform four such
reference changes we find $\phi=0$. For clarity, we can write down the
relative amplitudes of the many-body states $|\Psi_i\rangle$, where
$i$ labels the positions of the monomer:
\begin{eqnarray}
\Psi_1(D_a) &=& \Psi_2(D_b)=-\Psi_2(D_{b'})=-\Psi_4(D_c)=\Psi_4(D_{c'})\nonumber \\
			&=& \Psi_3(D_d)=-\Psi_3(D_{d'})=-\Psi_1(D_e)=\Psi_1(D_{e'}),\nonumber
\end{eqnarray} and we can see that in fact each reference change reverses the sign of the amplitude of the many-body state.

By combining the two results $\theta_{ex}=\pi$ and $\phi=0$, we
conclude that monomers have a statistical angle $\theta_s=\pi$, i.e.,
they obey Fermi statistics. This conclusion is completely independent
of how the two monomers exchanged. This is demonstrated with more
examples in Appendix~\ref{sec:addit-exch}, and can be proved for
general cases using the parton wave function introduced in the next
section.

\section{Parton construction.}
\label{sec:parton-construct}
\begin{figure}[htbp]
\centering
\includegraphics[width=.45\textwidth]{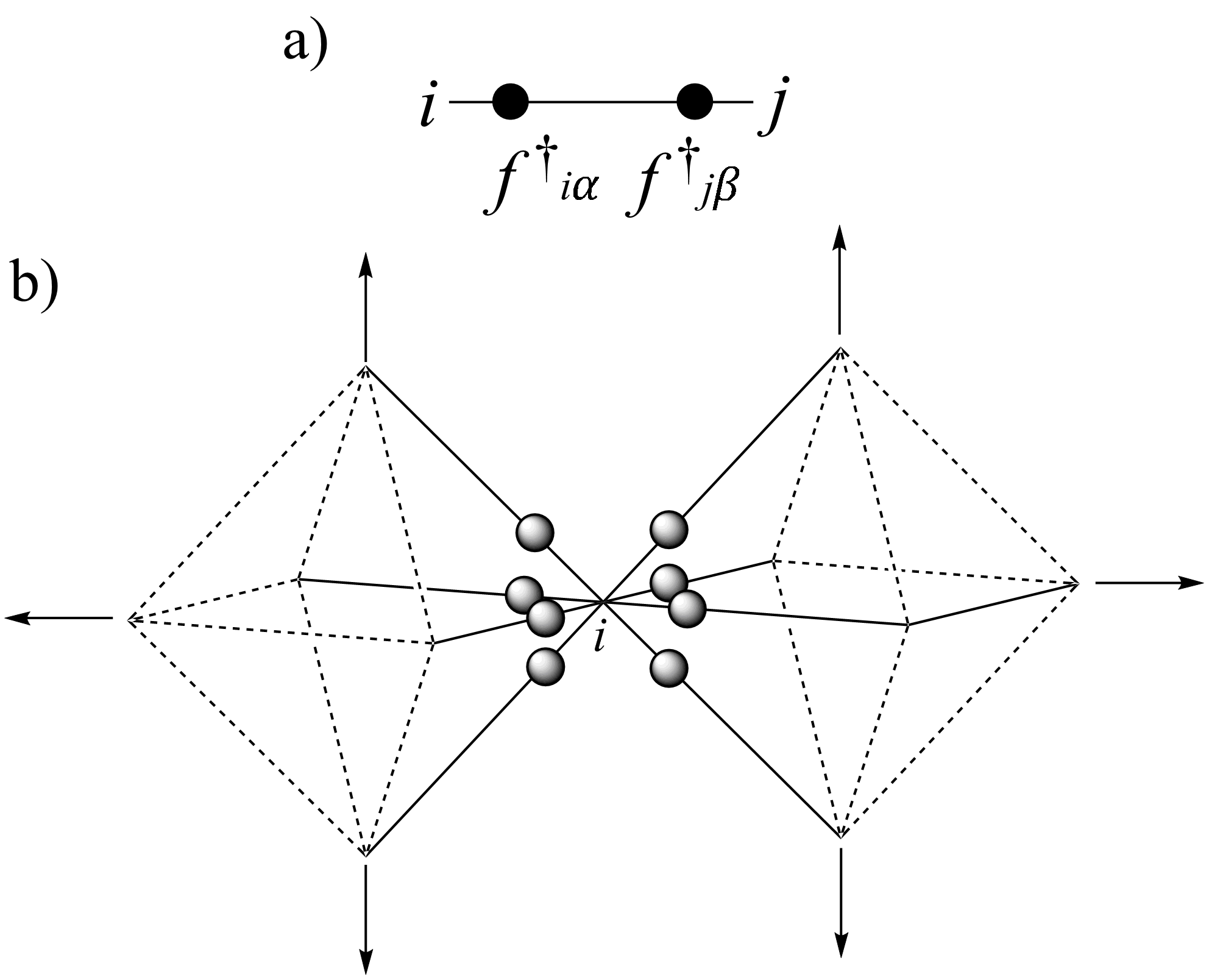}
\caption{Parton model. (a) One dimer degree of freedom is represented
  by two partons on the two ends of the dimer. (b) The partons live on
  a decorated CSO lattice in which every original CSO site is replaced
  by a eight-site cluster that represents the eight ways to draw a
  dimer from that site.}
\label{partonfig}
\end{figure}

In order to systemetically construct signful dimer
wave functions with fermionic monomer excitations, we present a
parton approach by writing a dimer degree of freedom in terms of two
fermion variables:
\beq
b^\dagger_{ij} = \xi_{ij} f^\dagger_{i \alpha} f^\dagger_{j \beta} , \; \; \; \xi_{ij} = - \xi_{ji} = \pm 1  \label{parton}
\eeq
where $b^\dagger_{i j}$ is a hardcore boson operator that creates a
dimer on a bond between two nearest-neighbor sites $i$ and $j$, and
the fermion operators $f^\dagger_{i \alpha}$ and $f^\dagger_{j \beta}$
are defined at two new sites, which are located near the two ends $i$
and $j$ of the bond $ij$, respectively (see
Fig.~\ref{partonfig}). These fermionic partons thus live on a
decorated CSO lattice made of a 8-site cluster surrounding every site
of the original lattice. The subscripts $\alpha$ and $\beta$
distinguish different sites within a cluster.

Due to the fermion anticommutation relation, we specify the ordering
of fermion operators on the right-hand side of (\ref{parton}) using
antisymmetric tensors $\xi_{ij}$ defined on all nearest-neighbor
bonds, each of which takes two possible values $\pm 1$.  Specifying
$\xi_{ij}$ on the bonds is equivalent to assigning the arrow pattern
as discussed previously, and doing it consistently with lattice
symmetries is central to our parton construction for quantum dimer
models.  For the CSO lattice here, we choose $\xi_{ij}$'s according to
the arrow pattern illustrated in Fig.~\ref{CSOarrows}: $\xi_{ij}=1$ if
an arrow starts at $i$ and ends at $j$, and $-1$ if it is the
opposite.  In essence, the parton construction (\ref{parton}) treats a
dimer as a composite of two fermions via the mapping between the
Hilbert space of dimers and that of fermions $|n\rangle_{ij}
\rightarrow |n\rangle_{i\alpha}\otimes |n\rangle_{j\beta}$, where
$n=0,1$ is the occupation number.  However, the Hilbert space of
fermions also includes unphysical states of the form
$|n\rangle_{i\alpha} \otimes |1-n\rangle_{j\beta}$, which have no
counterpart in the Hilbert space of dimers. Thus one must perform a
Gutzwiller projection that removes all unphysical configurations
within a parton wave function to generate a physical wave function.

A main advantage of the parton construction (\ref{parton}) is that
simple wave functions for partons can yield highly nontrivial wave
functions for dimers that are difficult to treat directly. Here we
consider a very simple parton wave function that can reproduce the
signful dimer wave function studied in
Sec.~\ref{sec:loops-sign-struct}, and we shall discuss other possible
extensions of parton wave functions as a way to construct variational
dimer wave functions in Sec.~\ref{sec:discussions}. In this section we
will focus on the following parton wave function \beq |\Phi_f\rangle
=\prod_i \frac{1}{\sqrt{8}} \sum_{\alpha=1}^{8} f^\dagger_{i \alpha} |
0 \rangle, \label{pt} \eeq which is a product of non-overlapping
``molecular'' orbitals, one per cluster.  For a given cluster, the
molecular orbital has equal amplitudes on all 8 sites.  We further
obtain physical dimer wave function $|\Psi \rangle$ from
$|\Phi_f\rangle$: \beq \Psi (i_1 j_1, ..., i_n j_n) = \langle 0 |
\left( \prod_{i_l \alpha_l, j_l \beta_l} \xi_{i_lj_l} f_{i_l \alpha_l}
  f_{j_l \beta_l} \right) | \Phi_f \rangle, \label{dimer} \eeq where
$\Psi (i_1 j_1, ..., i_n j_n)$ is the amplitude of having dimers on
nearest-neighbor bonds $i_1 j_1, ..., i_n j_n$.

The dimer wave function $|\Psi\rangle$ has several noteworthy
properties that can be inferred straightforwardly from the parent
parton wave function $|\Phi_f\rangle$. First, $|\Psi\rangle$ has
exactly one dimer touching every lattice site. This is because (i)
$|\Phi_f\rangle$ has exactly one parton per cluster, and (ii) two
partons in adjacent clusters that share a bond must be simultaneously
present to form a dimer.  Second, since partons have a uniform density
distribution in $|\Phi_f\rangle$, all possible dimer configurations
appear with equal probability in $|\Psi\rangle$, i.e., $| \Psi(i_1
j_1, ..., i_n j_n) | = \frac{1}{\sqrt{N_D}}$.  Last but most
importantly, due to the fermion nature of partons, the sign of
$\Psi(i_1 j_1, ..., i_n j_n)$ is not all positive but depends
nontrivially on the dimer coverings following the same sign rule as
stated in Eq.~(\ref{s}).
 
To obtain the nontrivial sign structure of $|\Psi\rangle$, we compare
the relative sign of $\Psi(D_1)$ and $\Psi(D_2)$, where $D_1$ and
$D_2$ are two arbitrary dimer configurations. We claim that the
relative sign is determined from the transition graph $G_{12}$ as the
following,
\begin{equation}
  \label{eo}
  \frac{\Psi(D_1)}{\Psi(D_2)}=\prod_m (-1)^{1+W(L_m) }=(-1)^{N(G_{12})+W(G_{12})},
\end{equation}
where $W(L_m)$ is the number of ``wrong-way'' arrows in the loop
$L_m$, and $N(G_{12})$ and $W(G_{12})$ are the same functions counting
number of loops and ``wrong-way'' arrows that were used in
Eq.~(\ref{s}). To prove (\ref{eo}), it suffices to consider a single
loop made of a sequence of lattice sites: $r_1 \rightarrow r_2
\rightarrow\cdots\rightarrow r_{2k} $. The two corresponding dimer
configurations are then $D_1=(r_1 r_2), \cdots, (r_{2k-1} r_{2k})$ and
$D_2 = (r_{2k} r_1), \cdots, (r_{2k-2} r_{2k-1})$.  It follows from
(\ref{pt}) and (\ref{dimer}) that $\Psi(D_1)$ and $\Psi(D_2)$ are
respectively given by
\begin{align*}
  \Psi(D_1) &= \xi_{r_1 r_2} \cdots \xi_{r_{2k-1} r_{2k}} \langle 0 | f_{r_1} f_{r_2} ... f_{r_{2k-1}} f_{r_{2k}} | \Phi_f \rangle \\
  \Psi(D_2) &= \xi_{r_{2k} r_1} \cdots \xi_{r_{2k-2} r_{2k-1}} \langle 0 | f_{r_{2k}}  f_{r_1} ...  f_{r_{2k-2}} f_{r_{2k-1}} | \Phi_f \rangle   
\end{align*}
The ratio is thus given by 
\begin{equation}
  \frac{\Psi(D_1)}{\Psi(D_2)} = -\prod_{i=1}^{2n} \xi_{r_i r_{i+1}}, 
  \; \; \; r_{2n+1} \equiv r_1. 
\end{equation}
where we have used the fermion anti-commutation relation. This proves
the sign rule (\ref{eo}).

Comparing the sign structures in Eq.~(\ref{s}) and Eq.~(\ref{eo}), one
can see that the wave function defined in Eq.~(\ref{psi}) for a certain
arrow configuration is the same as the parton wave function in
Eq.~(\ref{dimer}) using the corresponding $\xi_{ij}$
assignment. Therefore the parton approach gives a systematic way to
construct dimer wave functions with relative signs between different
dimer configurations. This construction also provides an explicit
proof of Eq.~(\ref{eq:smul}).

Furthermore, the parton approach gives a straightforward way to demonstrate
that the monomer excitations carry Fermi statistics. To see this, we
again study the phase difference between exchanging two holes and
moving one hole along the same path as we did in the previous section,
with the help of the reference parton wave function. For each step, a
monomer is moved by flipping one dimer. Particularly, assume that a
monomer is moved from site $i$ to site $j$ by moving one dimer from
$(jk)$ to $(ik)$. Correspondingly, in the parton wave function one
moves a hole with the following fermion hopping term,
\begin{equation}
  \label{eq:hhop}
  H_T^f = \sum_{ij}t_{ij}^ff_j f_i^\dagger,
\end{equation}
where $f_i=\frac1{\sqrt8}\sum_{\alpha=1}^8 f_{i\alpha}$, and
$t_{ij}^f=\pm1$ determines the relative sign between the wave functions
before and after hopping. Here $t_{ij}^f$ is chosen such that the
fermion hopping term $H_T^f$ has the same sign as the dimer hopping
term $H_T$ defined in Eq.~\eqref{eq:HT}. According to the relation
between dimer wave function and parton wave function in
Eq.~(\ref{dimer}), the relative phase of the physical wave function
before and after the projection is
\begin{equation}
  \label{eq:relph}
  \frac{\Psi((ik)\cdots)}{\Psi((jk)\cdots)}
  =\frac{\langle0|
    \xi_{ik}f_{i\alpha}f_{k\delta}t_{ij}^ff_jf_i^\dagger f_k^\dagger f_j^\dagger|0\rangle}
  {\langle0|
  \xi_{jk}f_{j\beta}f_{k\gamma} f_k^\dagger f_j^\dagger|0\rangle}
=\xi_{ik}\xi_{kj}t_{ij}^f.
\end{equation}
Here the left-hand side of the equation has the same sign as the
corresponding term in Eq.~\eqref{eq:tij}. Hence we choose the sign of
$t_{ij}^f$ such that the left-hand side has the sign of $t_{ij}$. This
is achieved by choosing
\begin{equation}
  \label{eq:tijf}
  t_{ij}^f = \xi_{ik}\xi_{kj}\sgn{t_{ij}}.
\end{equation}
With the condition in Eq.~(\ref{eq:tijf}) satisfied, after each step
of monomer hopping and dimer resonance, the sign of the dimer
wave function is always determined from the corresponding parton
wave function using the projection in Eq.~(\ref{dimer}). Hence the
monomer excitation in the dimer wave function has the same Fermi
statistics as the hole in the parton wave function.

It is important to note that the fermion hopping term in the parton
construction only depends on the beginning and ending sites $i$ and
$j$, while the monomer hopping from $i$ to $j$ is meditated by a dimer
move $(jk) \rightarrow (ik)$. For lattices with high symmetry, a
monomer hop from $i, j$ can be assisted by more than one such dimer
movers that are symmetry-related to each other, as shown in
Fig.~\ref{fig:exchange} for the CSO lattice. In this case, it is
crucial that the coefficients of $t_{ij}^f$ determined by
Eq.~(\ref{eq:tijf}) do not depend on the choice of possible dimer
moves specified by $k$.  We find that this is indeed the case for the
CSO lattice, thanks to the arrow pattern specified in
Fig.~\ref{CSOarrows}. This is because of the aforementioned fact that
any length-four loop has an even number of wrong-way arrows. In other
words, for any four points on a loop $i\rightarrow k\rightarrow
j\rightarrow k^\prime\rightarrow i$, we have
$\xi_{ik}\xi_{kj}=\xi_{ik^\prime}\xi_{k^\prime j}$, and therefore they
give the same $t_{ij}^f$. This also implies that with a given
$t_{ij}^f$, different ways of exchanging the monomer and a dimer
contribute to $t_{ij}$ with the same sign. Hence this provides a
general proof of this claim that was first introduced in
Sec.~\ref{sec:stat-mono}. Here our argument depends only on the fact
that the arrow pattern has an even number of wrong-way arrows for all
length-four loops, and can be generalized to other lattices where such
arrow pattern can be assigned.

\section{Discussions.}
\label{sec:discussions}

Using the CSO lattice as an example, we study a class of 3D quantum
dimer models with frustrated resonance terms. We construct an exact
ground state wave function of the model as a superposition of dimer
configurations with a twisted sign structure. The monomer excitations
in such a state are deconfined fermions. Furthermore, we give a
systematic approach to construct such states using parton projective
wave functions, and such construction naturally explains the Fermi
statistics of the monomer excitations.

The dimer model studied in this work can potentially be realized in
spin systems in a short-range RVB state where the spin-triplet gap is
much larger than the spin-singlet gap\cite{rk}. In such a state, the
spins are paired into spin-singlet valence bonds, and the ground state
is a superposition of different valence-bond configurations. Such a
state can be mapped to a dimer model by mapping spin valence bonds to
dimers. In the RVB state, a valence bond can be broken and a
spin-triplet excitation is created, similarly to the way a dimer
breaks up into two monomers in a quantum dimer model. Therefore the
monomer excitations in the dimer model can be viewed as spinon
excitations in the RVB state. The difference between a monomer and a
spinon, that the former carries a $U(1)$ quantum number but the latter
carries an $SU(2)$ quantum number, can be eliminated by applying a
Zeeman field to the RVB state. Therefore our signful wave function
(\ref{psi}) can be used to describe a class of 3D spin liquids with
fermionic spinon excitations. It is interesting to note that
antiferromagnetic interactions between spins naturally lead to {\it
  positive} resonance terms in quantum dimer models studied in this
work\cite{rk}.

One can extend our wave function to study dimer models with finite
density of monomers, which can be induced by applying a Zeeman field
larger than the spin-triplet gap to the corresponding RVB
state. Because monomers are fermions, the resulting state is likely to
realize a Bose metal state\cite{bm}, in which monomers form a Fermi
sea and boson correlation functions exhibit Fermi-surface-like
singularities. Using the parton construction presented in
Sec.~\ref{sec:parton-construct}, one can construct a variational
wave function for the doped dimer models by projecting a parton
wave function with a parton Fermi surface. Such wave functions can be
studied numerically using the variational Monte Carlo
method\cite{GROS1989}.

One can further regard the above RVB state as a description of an
electronic system at half filling.  Such a system will host spinless
holon excitations in addition to chargeless spinons. The Fermi
statistics of spinons then implies Bose statistics of holons. Doping
this system away from half filling can induce holon condensation and
leads to superconductivity\cite{rk, anderson}.  However, unlike the
original quantum dimer model, the superconducting state of a doped
quantum dimer model studied in this work will have $d$-wave pairing
symmetry. We shall leave these interesting extensions of our model to
future studies.

\begin{acknowledgments}
  We thank Senthil Todadri and Jeongwan Haah for invaluable
  discussions.  Y.Q. is supported by NSFC Grant No. 11104154. L.F. is
  supported by the DOE Office of Basic Energy Sciences, Division of
  Materials Sciences and Engineering under Award No. DE-SC0010526. We
  thank the hospitality of Tsinghua University during the LT26
  Satellite conference ``Topological Insulators and Superconductors'',
  when this work was initiated.
\end{acknowledgments}

\appendix

\section{Proof of multiplicity of transition graph signs under
  composition.}
\label{sec:addit-exampl-signs}

In this appendix we discuss the relation between the signs of
transition graphs and the sign of their composition, as described in
Eq.~\eqref{eq:smul}. In the main text we see that this equation holds
when loops in $G_{ab}$ and $G_{bc}$ do not overlap or overlap on one
bond. In this appendix we give a general proof for this
equation. Before going through the proof we first note that according
to the definition of $s(G_{ab})$ in Eq.~(\ref{s}), length-two loops
which are just two overlapping dimers always contribute a factor of
+1, so we can ignore them and only count non-trivial loops in
Eq.~(\ref{s}).

When two transition graphs $G_{ab}$ and $G_{bc}$ are composited
together, the transition loops may overlap on several pieces of
overlapping boundaries. The resulting graph $G_{ac}$ is obtained by
merging the loops at these overlapping boundaries. To prove
Eq.~(\ref{eq:smul}) it is sufficient to show that the sign of the
transition graph does not change before and after merging one piece of
overlapping boundary.

\begin{figure}[htbp]
  \centering
  \includegraphics{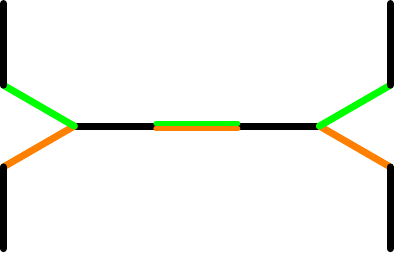}
  \caption{Overlapping boundary of two transition graphs. The black
    bonds belong to dimer covering $D_b$, green and orange bonds
    belong to dimer coverings $D_a$ and $D_c$, respectively, and the
    bond painted in both green and orange belongs to both $D_a$ and
    $D_c$.}
  \label{fig:bdry-odd}
\end{figure}

First, we argue that each piece of overlapping boundary contains an
odd number of bonds. As shown in Fig.~\ref{fig:bdry-odd}, a piece of
overlapping boundary must start and end on a bond in dimer covering
$D_b$, since for any bond on the boundary that belongs to coverings
$D_a$ and $D_c$, the bond adjacent to it belongs to $D_b$ and is thus
present in both $G_{ab}$ and $G_{bc}$ and also belongs to the
overlapping boundary. Between the two endings that belong to $D_b$,
the boundary consists of alternating bonds from $D_a/D_c$ and from $D_b$
respectively. Therefore the total number of bonds in an overlapping
boundary must be odd.

\begin{figure}[htbp]
  \centering
  \includegraphics{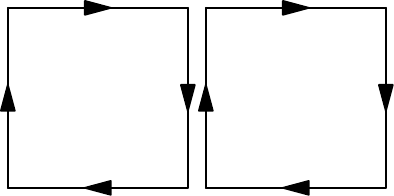}
  \caption{Two loops with one overlapping boundary. The orientations
    of the two loops are chosen such that the boundaries are oriented in
    opposite directions on the two loops, and the merged loop can
    naturally inherit the orientations.}
  \label{fig:merge2}
\end{figure}

When merging a piece of overlapping boundary between two different
loops, the two loops can be oriented such that they travel through the
overlapping boundary in opposite directions (see
Fig.~\ref{fig:merge2}), and after the merging the two loops are
combined into one that inherits the orientation of the two original
loops. After this merging the total number of loops is reduced by one,
and the number of wrong-way arrows is reduced by the length of the
boundary, which is an odd number, since each bond in the boundary is
traversed twice in two different directions in the original transition
graphs. Consequently according to the sign rule in Eq.~(\ref{s}) the
sign of the graph does not change before and after the merging.

\begin{figure}[htbp]
  \subfigure[\label{fig:merge1:1}]{\includegraphics[trim=0 0 0 -5mm]{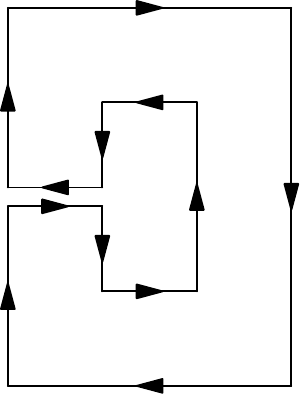}}
  \subfigure[\label{fig:merge1:1t}]{\includegraphics[trim=0 0 0 -5mm]{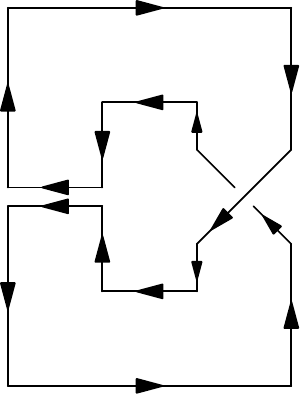}}
  \subfigure[\label{fig:merge1:1t2}]{\includegraphics{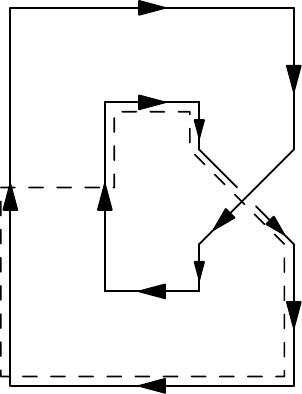}}
  \caption{One loop with one overlapping boundary. The arrow shows the
    orientation of the loop. (a) A planar loop without twisting. (b) A
    nonplanar loop with a twist. The loop is the edge of a M\"obius
    strip. (c) The result of merging of the loop in (b). Comparing to
    the orientation in (b), the direction of the bonds labeled by the
    dashed line are flipped.}
  \label{fig:merge1}
\end{figure}

On the other hand, when the piece of overlapping boundary belongs to
the same loop (this happens when two loops overlap at more than one
places: when merging the first piece of overlapping boundary it is the
case shown in Fig.~\ref{fig:merge2}, and all subsequential merges are
merging of the same loop), there are two possibilities: first, if the
loop is planar, as shown in Fig.~\ref{fig:merge1:1}, after the merging
the loop becomes two loops, and the number of loops is increased by
one. Similarly to the previous case, one can arbitrarily fix the
orientation of the loop and the overlapping boundary is traversed
twice in opposite directions. Hence the number of wrong-way arrows is
again reduced by an odd number. So the total sign stays the same.

The second case, which only exists in dimensions higher than two, is
when the loop is twisted and forms the boundary of a M\"obius strip, as
shown in Fig.~\ref{fig:merge1:1t}. In this case after the merging
there is still one loop. Moreover, if we fix the orientations of the
old and the new loop, the orientation needs to be flipped on one
portion of the loop at the merging. The portion where the orientation
is flipped, labeled by the dashed line in Fig.~\ref{fig:merge1:1t2},
also forms a transition loop itself and therefore contains an even number
of bonds. Consequently the number of wrong-way arrows is changed by an
even number. Combined with the invariant loop count this implies that
the sign stays the same.

\section{Calculations of monomer statistics for various exchanges.}
\label{sec:addit-exch}

In this appendix we will further demonstrate the Fermi statistics of
monomers by calculating the phase change $\theta_{ex}$ and background
flux $\phi$ for various exchanges. We will determine these signs by
considering the transition loops necessary to perform reference
changes during each exchange using the procedure developed in
Sec.~\ref{sec:stat-mono}.

\begin{figure}[htbp]
\centering
\includegraphics[width=.45\textwidth]{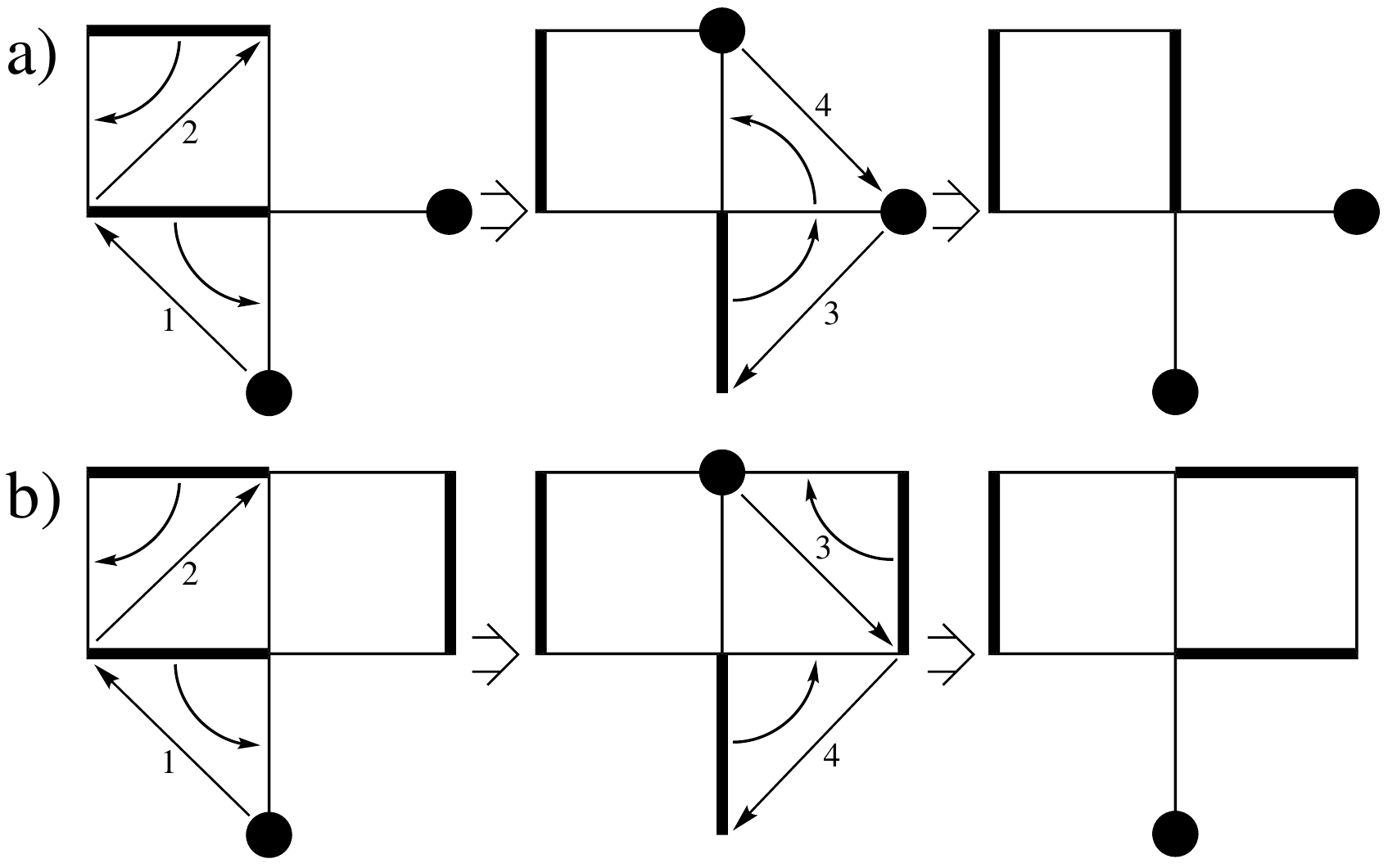}
\caption{(a) The two holes are exchanged on the square sublattice
  using next-nearest-neighbor hops, which requires only one square
  loop. (b) We move a single hole along the same path to find the
  background flux. Two square loops are needed, giving $\phi=0$. }
\label{NNNexchange}
\end{figure}

In Sec.~\ref{sec:stat-mono}, we showed an exchange which only used
nearest-neighbor hops. We will now exchange two holes using only
next-nearest-neighbor hops on the square lattice
[Fig.~\ref{NNNexchange}(a)], which can be obtained by taking the
subset of lattice sites that lie on a plane bisecting a layer of CSO
unit cells. This exchange requires only one change of reference at the
very end to return to the initial dimer tiling, which uses a
transition graph with a single square loop, yielding $\theta_{ex} =
\pi$. In order to calculate the statistical angle $\theta_s$, we need
to find the flux $\phi$ for a single hole moving around the same path,
using a different ordering of the hopping terms, along with new
corresponding reference changes [Fig.~\ref{NNNexchange}(b)]. This time,
there are two reference changes, each using a square loop. Since
hopping terms do not affect the relative signs of the many-body states
in the process, we only need to find the effect of the reference
changes on the sign. For a general process with many different
reference changes, we calculate the signs of each loop involved
individually using Eq.~(\ref{s}), and then determine the overall sign
by taking the product of all the individual signs of the loop using
Eq.~(\ref{eq:smul}). For this particular case, two square loops give
us an overall sign of $+1$, meaning $\phi = 0$. Therefore, $\theta_s =
\theta_{ex} - \phi = \pi$.

\begin{figure}[htbp]
\centering
\includegraphics[width=.45\textwidth]{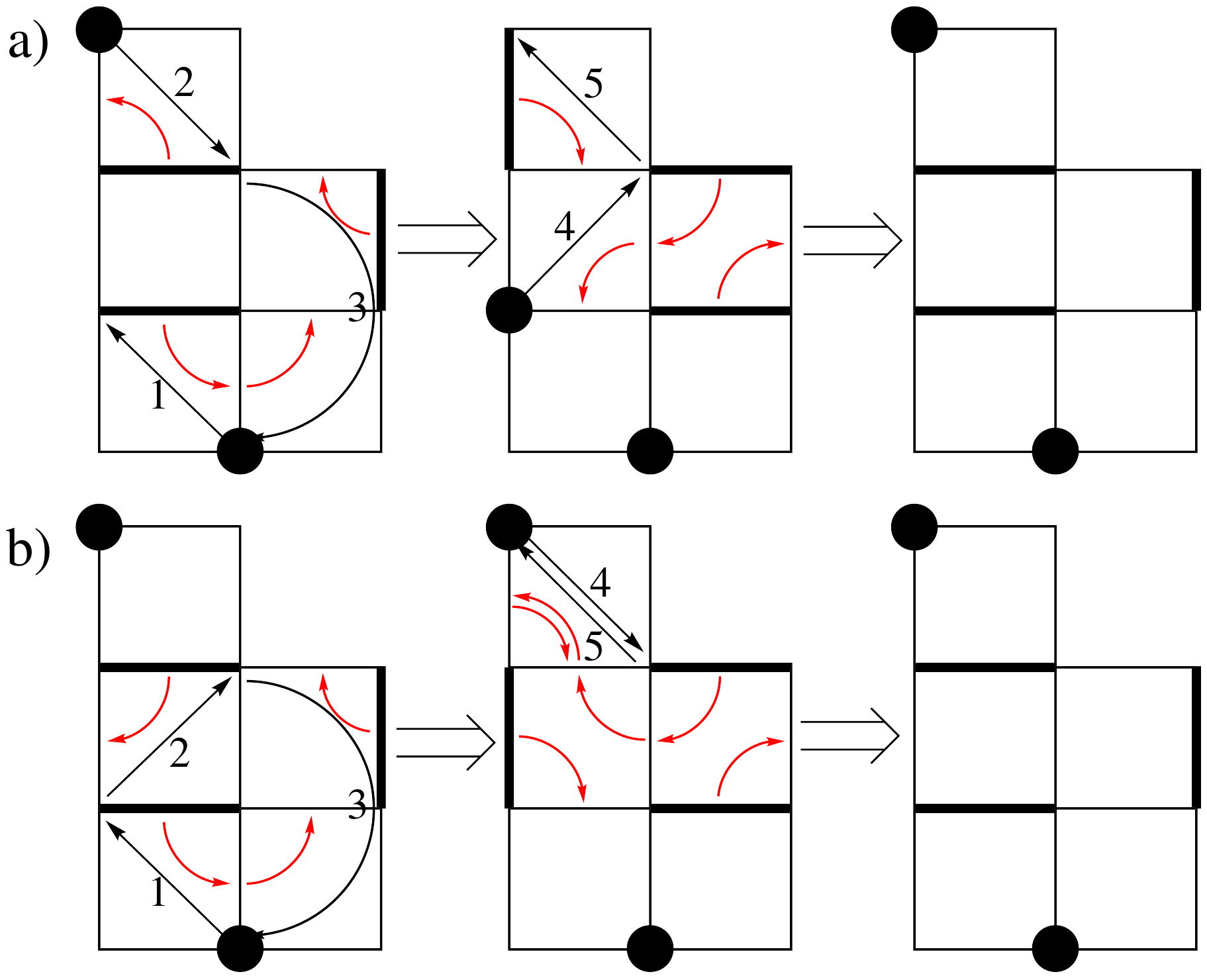}
\caption{(a) A Levin-Wen exchange is performed on two holes. (b) We
  move a single hole along the same path to find that background flux
  $\phi$ = 0. }
\label{SquareLW}
\end{figure}

We can perform a similar calculation using the five-step Levin-Wen
exchange described in Ref.~\onlinecite{lw}. This approach sometimes
has the advantage of producing more convenient exchanges, and can be
used to calculate both $\theta_{ex}$ and $\phi$ using the same two
monomers by performing a process of exchange and nonexchange,
respectively, which will depend on the order of hopping terms
(Fig.~\ref{SquareLW}). In this case, the exchange process uses a
single square loop, while the nonexchange process uses two square
loops, which gives $ \theta_{ex} = \pi, \phi = 0, \theta_s = \pi$.

\begin{figure}[htbp]
\centering
\includegraphics[width=.45\textwidth]{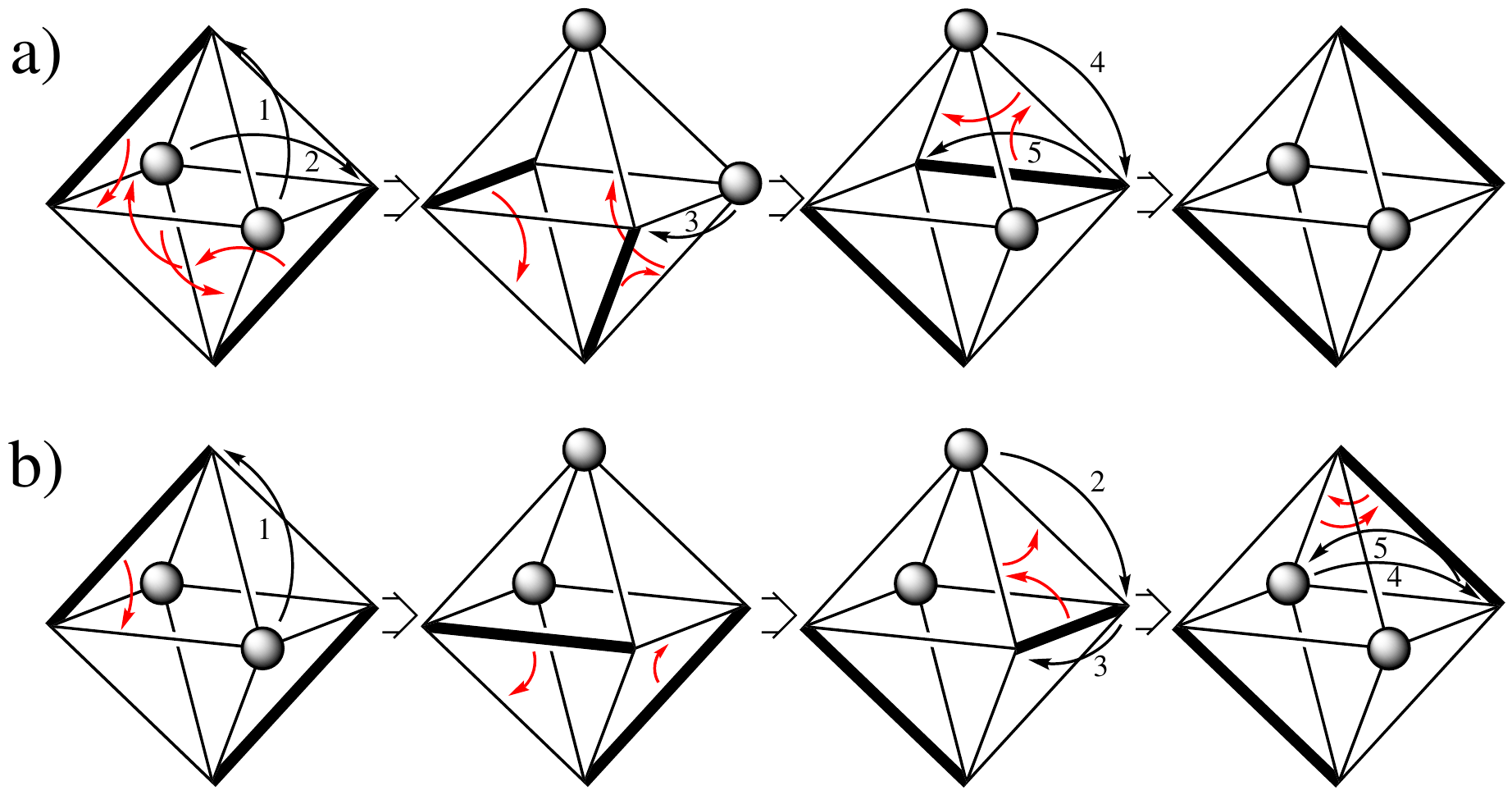}
\caption{(a) A Levin-Wen exchange is performed on two holes. Two reference changes occur after the second and third hops. (b) The sequence of hops on two holes are repeated, this time without exchange. Only a single reference change is necessary, after the first hop.}
\label{LW}
\end{figure}

For completeness, we now present a similar Levin-Wen exchange using
only nearest-neighbor hops (Fig.~\ref{LW}). Here we change reference
twice, using a bent-square loop each time, while the initial and final
dimer tilings are connected by a single square loop. Therefore the
exchange angle is $\theta_{ex} = \pi$. For the nonexchange process
, we need one bent-square loop for a change of reference, and one
square loop to connect initial and final dimer tilings, which give a
background flux of $\phi$ = 0, so $\theta_s = \pi$.

\begin{figure}[htbp]
\centering
\includegraphics[width=.45\textwidth]{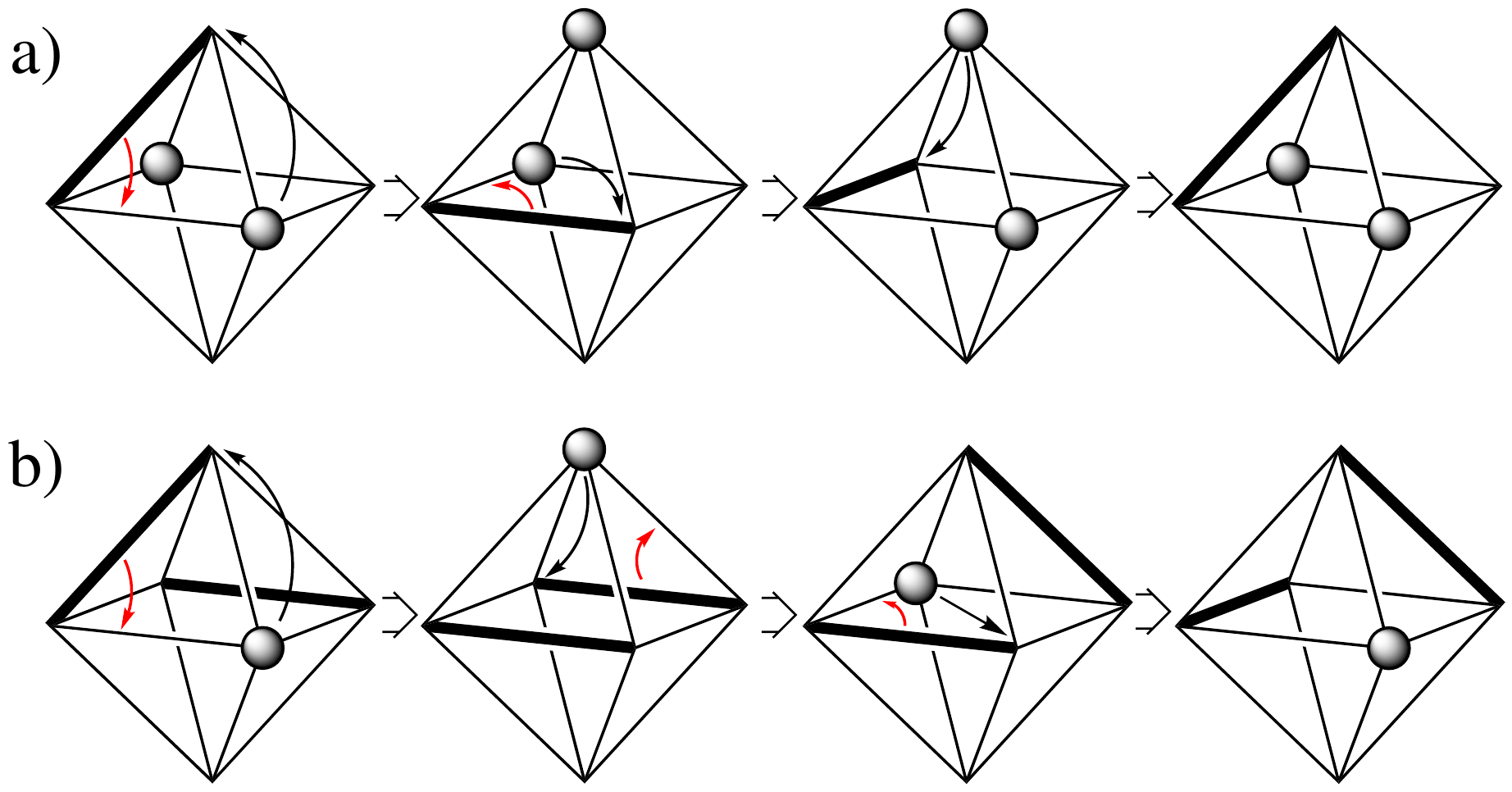}
\caption{(a) Two holes are exchanged using two nearest-neighbor hops, and one next-nearest neighbor hop. Red arrows show corresponding dimer moves. (b) A single hole is moved along the same path, generating non-zero flux.}
\label{W1}
\end{figure}

So far, we have considered exchanges which have used only one type of
hole-hopping term, and all exchanges have zero background flux. We now
present an exchange which uses both nearest-neighbor and
next-nearest-neighbor hopping and has nonzero background flux. It
also has the interesting property that the exchange itself does not
change the reference configuration. It is trivial to calculate that
$\theta_{ex} = 0$, since initial and final dimer tilings are
identical, and no reference changes are needed (Fig.~\ref{W1}). When
moving a single hole on the same path we require one bent-square loop,
meaning that the background flux is nonzero: $\phi = \pi$. In this
case, however, the exchange angle was 0, so the statistical angle
$\theta_s$ is still $\pi$.

These various examples demonstrate the persistence of Fermi statistics
of monomer holes on the CSO lattice, regardless of the type of hopping
terms included in the Hamiltonian, or the presence of background flux.


\begin{thebibliography}{10}
\bibitem{qmcnote} The model with $J_i>0$ can be studied with the
  quantum Monte Carlo method without the sign problem after a nonlocal
  unitary transformation that changes the sign of the dimer basis
  using the factor $s(G)$ defined in Eq.~\eqref{s}, which maps the
  model to the one with $J_i<0$.
\bibitem{monomernote}Since monomers are fractionalized excitations,
  there must be an even number of monomers to form a valid dimer
  configuration, and here in the single-monomer wavefunction
  $|\Psi_i\rangle$ we assume there is another monomer fixed at a
  location far away from $i$.
\end{thebibliography}
\end{document}